\documentclass[fleqn,10pt]{STIM2c} 

\definecolor{color1}{RGB}{0,0,90} 
\definecolor{color2}{RGB}{0,20,20} 

\newlength{\tocsep} 
\usepackage{color}
\usepackage[binary-units=true]{siunitx}
\usepackage{chemformula}
\usepackage{lettrine}
\usepackage{lipsum} 
\usepackage{setspace}
\usepackage{booktabs}

\usepackage[dvipsnames]{xcolor}

\JournalInfo{Preprint for Review, 2019} 

\PaperTitle{Three-Dimensional Microscopy by Milling with Ultraviolet Excitation} 

\Authors{Jiaming Guo\textsuperscript{1}, Camille Artur\textsuperscript{2}, Jason L. Eriksen\textsuperscript{3}, and David Mayerich\textsuperscript{1,2}*} 
\affiliation{\textsuperscript{1}\textit{University of Houston}, Department of Electrical and Computer Engineering} 
\affiliation{\textsuperscript{2}\textit{University of Houston}, NSF BRAIN Center}
\affiliation{\textsuperscript{3}\textit{University of Houston}, Department of Pharmacology} 
\affiliation{*\textbf{Corresponding author}: mayerich@uh.edu} 

\newcommand{\keywordname}{Keywords} 

\Abstract{Analysis of three-dimensional biological samples is critical to understanding tissue function and the mechanisms of disease. Many chronic conditions, like neurodegenerative diseases and cancers, correlate with complex tissue changes that are difficult to explore using two-dimensional histology. While three-dimensional techniques such as confocal and light-sheet microscopy are well-established, they are time consuming, require expensive instrumentation, and are limited to small tissue volumes. Three-dimensional microscopy is therefore impractical in clinical settings and often limited to core facilities at major research institutions. There would be a tremendous benefit to providing clinicians and researchers with the ability to routinely image large three-dimensional tissue volumes at cellular resolution. In this paper, we propose an imaging methodology that enables fast and inexpensive three-dimensional imaging that can be readily integrated into current histology pipelines. This method relies on block-face imaging of paraffin-embedded samples using deep-ultraviolet excitation. The imaged surface is then ablated to reveal the next tissue section for imaging. The final image stack is then aligned and reconstructed to provide tissue models that exceed the depth and resolution achievable with modern three-dimensional imaging systems.}

\begin{document}

\flushbottom 

\maketitle 


\thispagestyle{empty} 



\addcontentsline{toc}{section}{\hspace*{-\tocsep}Introduction} 
Current biomedical research and clinical diagnoses rely heavily on histological tissue sectioning to visualize phenotype and phenotypic changes. However, two-dimensional sectioning provides a very limited representation of three-dimensional structure. These limitations are particularly difficult to reconcile for complex three-dimensional structures, such as neural and microvascular networks. Cancer and neurodegenerative diseases affect the surrounding tissue phenotype, introducing complex changes in tissue structure. For example, tumor-induced vascular endothelial growth factor (VEGF) stimulates angiogenesis, providing necessary substrates for tumor cell growth and spreading \cite{goel_vegf_2013}. Neurodegenerative disorders such as Alzheimer's disease (AD) induce currently irreversible damage in vasculature and neural connectivity \cite{bell_neurovascular_2009}. These changes are extremely difficult to quantify with traditional histological sections, which are only \numrange{4}{6}\si{\micro\meter} thick.

Three-dimensional (3D) microscopy, such as confocal, multi-photon, and light-sheet microscopy are common methods of 3D imaging. However, these techniques are extremely time-consuming, limited to small ($\approx$\SI{1}{\milli\metre} thick) samples \cite{helmchen_deep_2005}, and often expensive. Several attempts have been made to alleviate constraints on sample thickness, including optical modifications \cite{vettenburg_light_sheet_2014} and tissue clearing \cite{rocha_tissue_2019}. In addition to trade-offs in resolution, imaging depth is still limited by objective working distance and speed is limited by photobleaching. Physical sectioning overcomes thickness constraints, with advanced developments including vaporizing imaged layers \cite{tsai_all-optical_2003} and physical cutting \cite{mayerich_knife_edge_2008,ragan2012serial}. However, these methods are still either time-consuming or cost prohibitive, making them impractical for most research and clinical settings.

\begin{figure}[tb]
  \centering
  \includegraphics[width=\the\columnwidth]{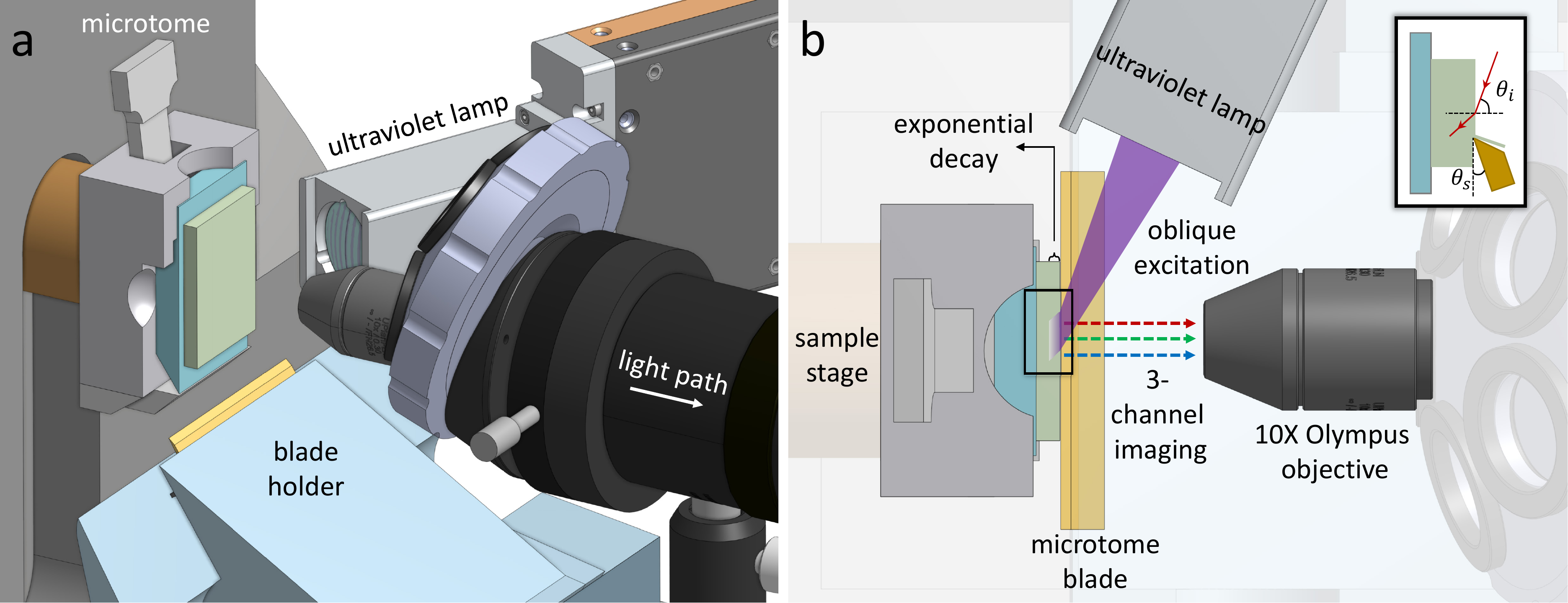}
  \caption{MUVE imaging. (a) Side view of the MUVE instrumentation showing (1) an automated microtome, (2) deep-ultraviolet (\SI{280}{\nano\meter}) source, and (3) standard microscope objective. (b) A planar view shows the optical train, where UV light incident on the sample excites fluorescent labels that are collected by the microscope objective. After imaging, a microtome ablates the tissue section and re-positions the sample for imaging the next section.}
  \label{fgr:microtome}
\end{figure}

In this paper, we propose a new methodology that offers (1) fast acquisition speeds comparable to 2D histology, (2) unlimited sample thickness, (3) resolution that exceeds the diffraction limit along the axial direction, and (4) simple and low-cost construction. This imaging system is based on recent innovations in deep-ultraviolet histology \cite{fereidouni_microscopy_2017} that allow block-face imaging of fresh samples to achieve histology-like 2D images. We describe the design of our prototype imaging system (Figure \ref{fgr:microtome}) and provide a demonstration and characterization of its performance for imaging high-resolution vascular and cellular components in formalin-fixed and paraffin-embedded (FFPE) samples.

\section{Theoretical Approach}

\subsection{Deep Ultraviolet Optical Sectioning}
\begin{figure*}[tb]
  \centering
  \includegraphics[width=\textwidth]{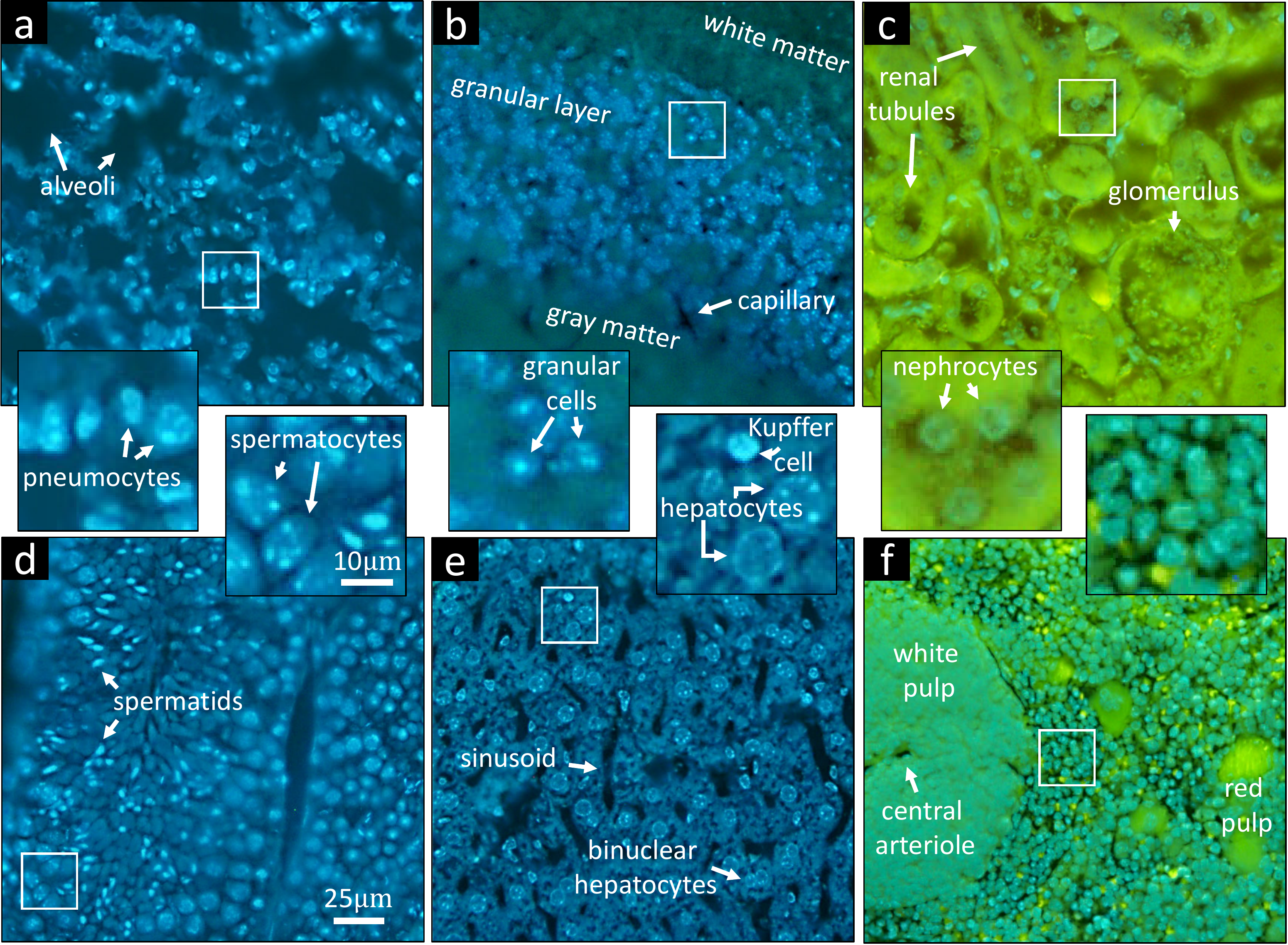}
  \caption{MUVE imaging of different mouse organs embedded in UV27-doped paraffin wax. (a) Singleplex imaging of mouse lung stained only with HO342. (b) Duplex imaging of mouse cerebellum perfused with India-ink and treated with DAPI. (c) Duplex imaging of mouse kidney stained with Eosin and HO342. (d) Singleplex imaging of mouse testicle stained only with HO342. (e) Duplex imaging of mouse liver perfused with India-ink and treated with HO342. (f) Duplex imaging of mouse spleen stained with Eosin and HO342.}
  \label{fgr:muse}
\end{figure*}

The proposed instrument is inspired by a new imaging technology known as \textit{microscopy with ultraviolet surface excitation} (MUSE) \cite{fereidouni_microscopy_2017} that allows slide-free histology on intact tissue using fluorescent dyes. The main advantage of UV excitation is that light penetration under direct illumination is limited to the sample surface (\SI{10}{\micro\meter} or less) \cite{meinhardt_wavelength_dependent_2008}. Many common fluorophores are excited by deep UV, including 4$^\prime$,6-diamidino-2-phenylindole (DAPI), Hoechst 33342 (HO342), and eosin (Figure \ref{fgr:muse}). Since glass optics block UV light, no excitation/emission filters or dichroic mirrors are necessary, significantly reducing the cost of optics while allowing simultaneous multi-channel imaging using a color camera. In addition, the necessary deep-UV optics are inexpensive and readily available in the form of quartz lenses.

Penetration depth can be further controlled by using a higher incident angle, however there is a trade-off with illumination intensity. Furthermore, recent research shows that water-immersion MUSE achieves approximately 50$\%$ reduction in imaging depth compared with air-immersion MUSE \cite{yoshitake_rapid_2018}.

\begin{figure*}[tb]
  \centering
  \includegraphics[width=\textwidth]{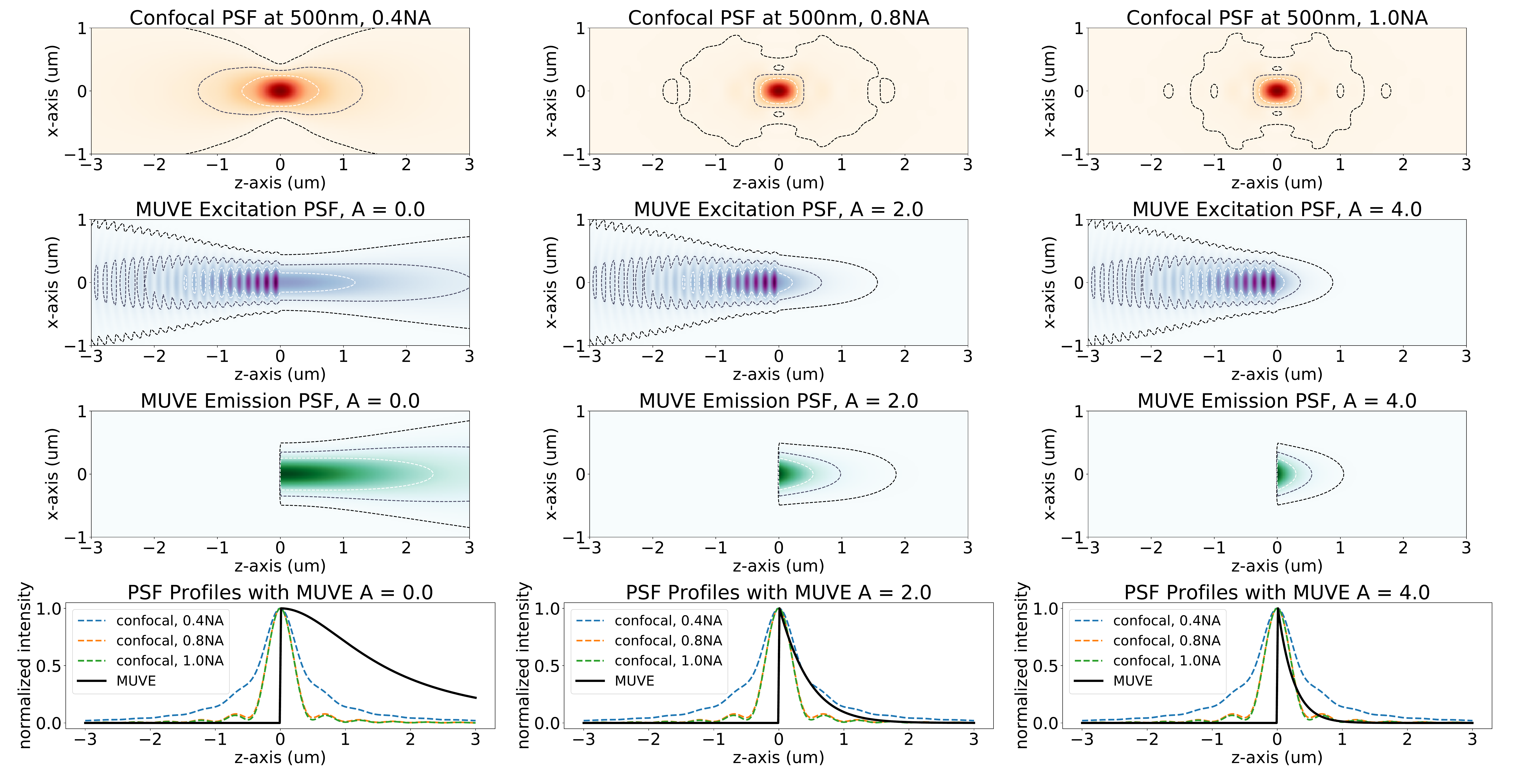}
  \caption{Monte-Carlo simulations of confocal and MUVE point-spread-functions using coupled-wave theory for absorbance in a layered homogeneous substrate \cite{davis2010theory}. All simulations show $x$-polarized coherent light propagating from left to right and intensities are normalized for each image. Contours indicate (from darkest to lightest) 1\%, 10\%, and 30\% thresholds of maximum intensity. (red) Confocal PSFs for imaging in idealized (i.e. cleared) samples using 0.4NA (left), 0.8NA (center), and 1.0NA (right) objectives. In MUVE imaging, exponential absorbance of the excitation is the dominant factor describing the axial PSF. (purple) Incident deep-UV light is shown incident on a sample using a low-NA ($\approx$0.25) objective. Varying the molar absorbance by doping the embedding compound reduces penetration, creating a smaller axial PSF. A UV-transparent sample (left) shows a significant contribution from back-scattered light. However, doping with UV27 (center, right) results in a significant improvement over high-end confocal imaging. (green) The excited region of the samples scales with the penetrating UV PSF, however ablation results in truncation of the left half. (bottom) A comparison is shown between confocal and MUVE axial PSFs. Lateral resolution is theoretically identical between MUVE and confocal.}
  \label{fgr:muve_simulations}
\end{figure*}

The proposed approach relies on doping the sample embedding medium with a soluble UV-absorbant dye (UV27, Epolin). Increasing the dye concentration within the embedding medium reduces the axial point-spread-function (PSF), providing optically thinner sections (Figure \ref{fgr:muve_simulations}).

\subsection{Serial Ablation 3D Imaging}
Several approaches have been proposed for integrating block-face microscopy with serial ablation to enhance axial resolution and image depth. Early studies rely on all-optical imaging and ablation \cite{tsai_all-optical_2003}, which is time-consuming but applicable to a wide range of tissues. Serial block face scanning electron microscopy (SBF-SEM) \cite{denk2004serial} uses a microtome blade for ablation, providing nanometer-scale resolution of samples embedded in hard polymers. Alternative approaches achieve similar results using focused ion beams \cite{knott2008serial}. However, these methods are limited to extremely small micrometer-scale samples and lack molecular specificity.

Integration of microtome sectioning with optical approaches has been proposed for large-scale imaging. However, these instruments are extremely expensive to construct and difficult to maintain. For example, knife-edge scanning microscopy (KESM) \cite{mayerich_knife_edge_2008,li2010micro} requires high-precision stages and time-consuming sample protocols, while two-photon tomography \cite{ragan2012serial} requires expensive two-photon imaging systems.

\begin{figure}[tb]
  \centering
  \includegraphics[width=\the\columnwidth]{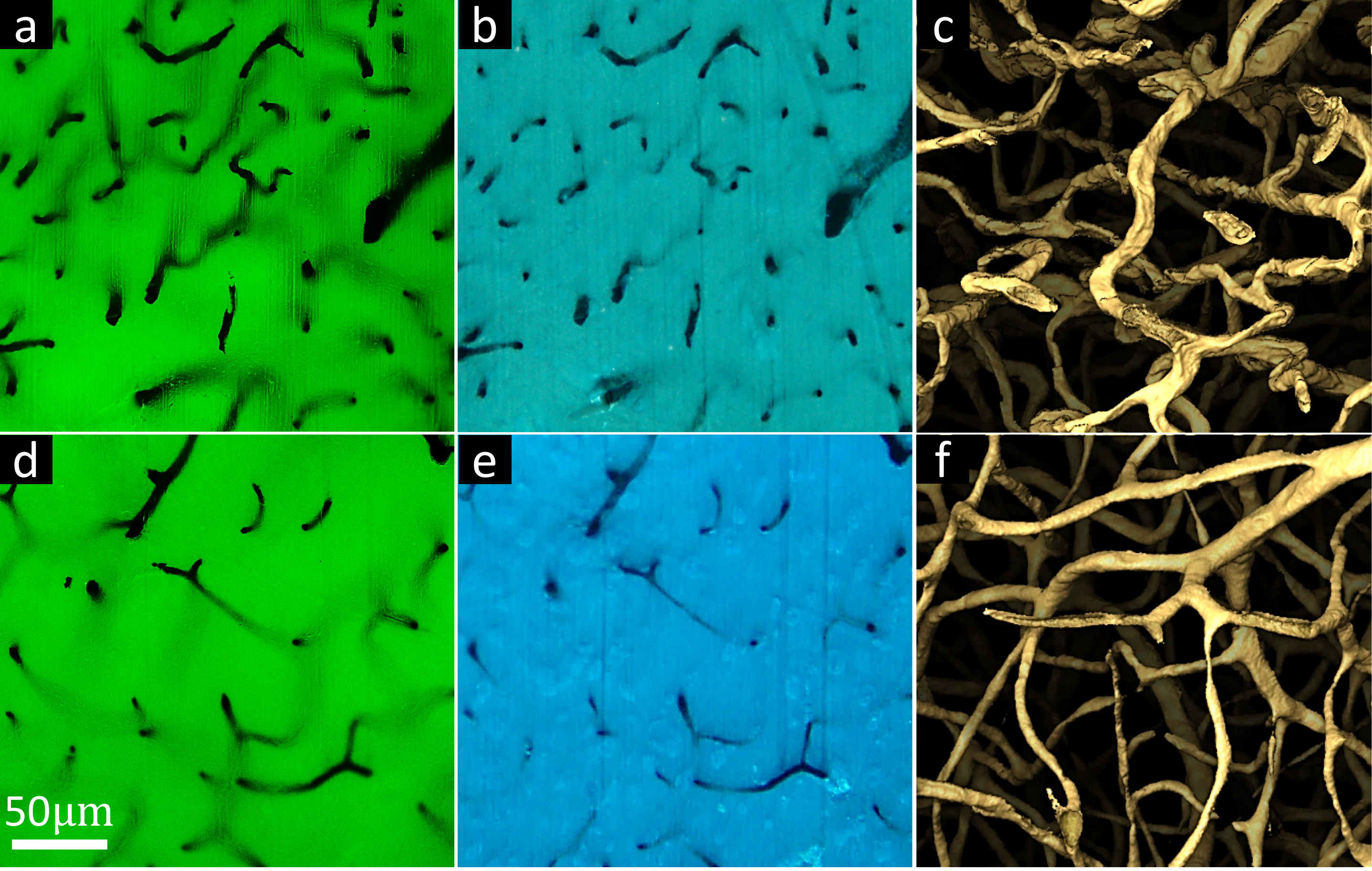}
  \caption{Advantages of MUVE imaging as 3D microscopy. (a) Block-face imaging of paraffin-embedded brain (top) and UV27-doped paraffin-embedded brain (bottom) using a wide-field fluorescence microscope with DAPI excitation (\SI{390}{\nano\metre}). Near-visible penetration in tissue is large in both cases, making it impossible to reconstruct the 3D structure of microvessels. (b) Block-face imaging of paraffin-embedded brain (top) and UV27-doped paraffin-embedded brain (bottom) using MUVE (same regions as shown in (a)). Deep-UV penetration in tissue is significantly shorter than that of near-visible and UV27 infiltration further reduces the excitation volume. (c) Isosurface rendering of paraffin-embedded brain (top) shows uneven vessel surface reconstruction whereas isosurface rendering of UV27-doped paraffin-embedded brain (bottom) shows sharp and smooth vessel surface reconstruction.}
  \label{fgr:BlockComparison}
\end{figure}

The proposed approach, which we refer to as MUSE milling or \textit{milling with ultraviolet excitation} (MUVE) relies on block-face imaging, requiring the attachment of common microscope optics to an automated microtome. By leveraging the proposed UV-blocking approach described above, it is possible to image formalin-fixed paraffin embedded (FFPE) samples using MUSE, while milling away imaged sections using the microtome. This allows high-resolution three-dimensional reconstruction of complex samples using both fluorescent (Figure \ref{fgr:muse}) and absorbing (Figure \ref{fgr:BlockComparison}) dyes.

\subsection{MUVE Instrumentation}
Our prototype MUVE imaging system (Figure \ref{fgr:microtome}) is based on an HM355S motorized microtome (Thermo Fisher Scientific) capable of automated \numrange{0.5}{100}\si{\micro\meter} sectioning. Our modifications include a FireEdge FE200 LED capable of up to \SI{300}{\milli\watt} emission centered at \SI{280}{\nano\meter} (Phoseon Technology, Hillsboro OR). Custom UV optics were designed to focus the UV source to a \SI{1}{\milli\meter} spot at the sample block face. A custom microscope is mounted laterally to observe the block face. The light path of this microscope is simply composed of a 10X objective (Olympus Plan Fluorite objective, 0.3NA), a tube lens (Olympus U-TLU), and a 0.5X camera adapter (Olympus U-TV0.5XC-3). A Thorlabs CNS500 objective turret is used to support additional objectives. Emitted fluorescence was detected using a line-scan color camera (Thorlabs 1501C-GE) that provides a theoretical throughput of 1392$\times$\SI{23}{\hertz}$\approx$\SI{32}{{\kilo pixel}\per\second} at 3 colors per pixel, resulting in a throughput of approximately \SI{96}{{\kilo\byte}\per\second}. This microscope was rigidly mounted to a two-axis translation stage (Thorlabs XYT1) for positioning and focusing.

\section{Materials and Methods}
\subsection{Tissue Collection and Labeling}
Mice were euthanized using \ch{CO2} based on guidelines provided by the American Veterinary Medical Association (AVMA). Mice were then perfused transcardially with \SI{20}{\milli\liter} of room temperature phosphate-buffered saline (PBS) solution (pH 7.4), followed by \SI{20}{\milli\liter} of room temperature 10\% neutral-buffered formalin (pH 7.4). Perfusion with PBS and formalin removes blood from the circulatory system and fixes the tissue.

Mice were then perfused with \SI{10}{\milli\liter} of undiluted India-ink at a rate of \SI{\approx 1}{{\milli\liter}\per\second}. We tested multiple vascular stains, including polyurethane resin (vasQtec PU4ii) and fluorescent tattoo ink (Skin Candy). Both fluorescent labels provided excellent contrast using block-face imaging. However, vasQtec resin was degraded by alcohol during dehydration prior to perfusion (both ethanol and isoproyl alcohol were tested). While the fluorescent tattoo inks survived embedding, the dyes were composed of fluorescent particles \SI{\approx 1}{\micro\meter}, resulting in blockages that prevented capillary labeling. We found that India ink (Higgins) \cite{mayerich2011fast} provided adequate perfusion and contrast for MUVE imaging.

Organs were then removed and fixed in 10\% neutral-buffered formalin for \SI{24}{h} and finally stored in 70\% ethanol (\ch{C2H5OH}). Optionally, tissue samples were also stained using a variety of compounds to provide cellular contrast, including DAPI, Hoechst, and eosin \cite{fereidouni_microscopy_2017} for fluorescent imaging and thionine \cite{mayerich_knife_edge_2008,choe2011specimen} for negative-contrast Nissl staining.

\subsection{Specimen Preparation and Embedding}
Organs were embedded in paraffin wax for imaging. UV penetration was controlled by doping molten paraffin with up to 14\% UV27 dye (Epolin). Similar protocols were followed for all ranges of doped paraffin infiltration. Organ sections were dehydrated through a series of graded ethanols (\numrange{70}{100}$\%$) over the course of \SI{8}{h}, followed by clearing with xylene substitute (SIGMA A5597) for \SI{3}{h}. Standard paraffin wax (Tissue-Tek Paraffin) was selectively doped with UV27 at \SI{60}{\celsius}, and samples were soaked in the selected mixture for \SI{2}{h} to allow infiltration. The paraffinization process was performed with the aid of a tissue processor (Leica TP1020). Note that tissue shrinkage is always expected during paraffinization procedures and the degree of shrinkage can reach up to 40$\%$ in volume for brain tissue. This can be potential avoided using matrices that have low shrinkage artifacts, such as glycol-methacrylate resins (Electron Microscopy Sciences Technovit 7100) or urethane rubbers (Smooth On Clear Flex 95). In particular, we found that Technovit was highly UV opaque, but significantly more difficult to mill.

Other nuclear stains, such as DAPI and Hoechst (HO342), are compatible with India ink perfusion. While these stains are subject to bleaching during paraffin infiltration, we have found that paraffinized samples can be stained with DAPI and Hoechst, with penetration up to \SI{1}{\milli\metre} after 3 days of in solution. For example, the Hoechst solution was prepared by diluting the HO342 stock solution (Thermo Fisher Hoechst 33342) 1:2000 in 1X PBS. This also allows staining of \numrange{1}{5}\si{\micro\metre} embedded tissue (Figure \ref{fgr:muse}). Staining was performed by covering the block face with solution for \numrange{2}{3}\si{\min} prior to imaging.

\subsection{Image Collection}
Conventional microtome blades (DURAEDGE Low Profile) were used for cutting, with a cutting angle of $10^{\circ}$ (Figure \ref{fgr:microtome}b). The single stroke operation mode of the microtome was used for semi-automated acquisition. Cutting velocities were randomized to prevent the reinforcement of artifacts such as knife chatter. However, the resting position of the microtome oscillates slightly around its central position causing an offset along the cutting direction. We applied an automated alignment algorithm available in OpenCV \cite{evangelidis_parametric_2008} for compensation. Camera triggering used an image acquisition software package (Thorlabs ThorCam) controlled using an external TTL signal. Images were acquired using \SI{100}{\milli\second} exposure with a digital gain of 40, and image corrections were performed to adjust brightness and contrast using ImageJ. It took approximately \SI{4}{\second} for each slice: \SI{2}{\second} for cutting, \SI{1}{\second} for stage stabilization, and \SI{1}{\second} for image capturing. This process takes approximately 2 hours to collect 2,000 slices. While this prototype system uses a commercial microtome, a fully-automated system could achieve a data rate similar to 3D color histology for three-dimensional samples.

\section{Results}
\subsection{Point Spread Function Characterization}
\begin{figure}[tb]
  \centering
  \includegraphics[width=\the\columnwidth]{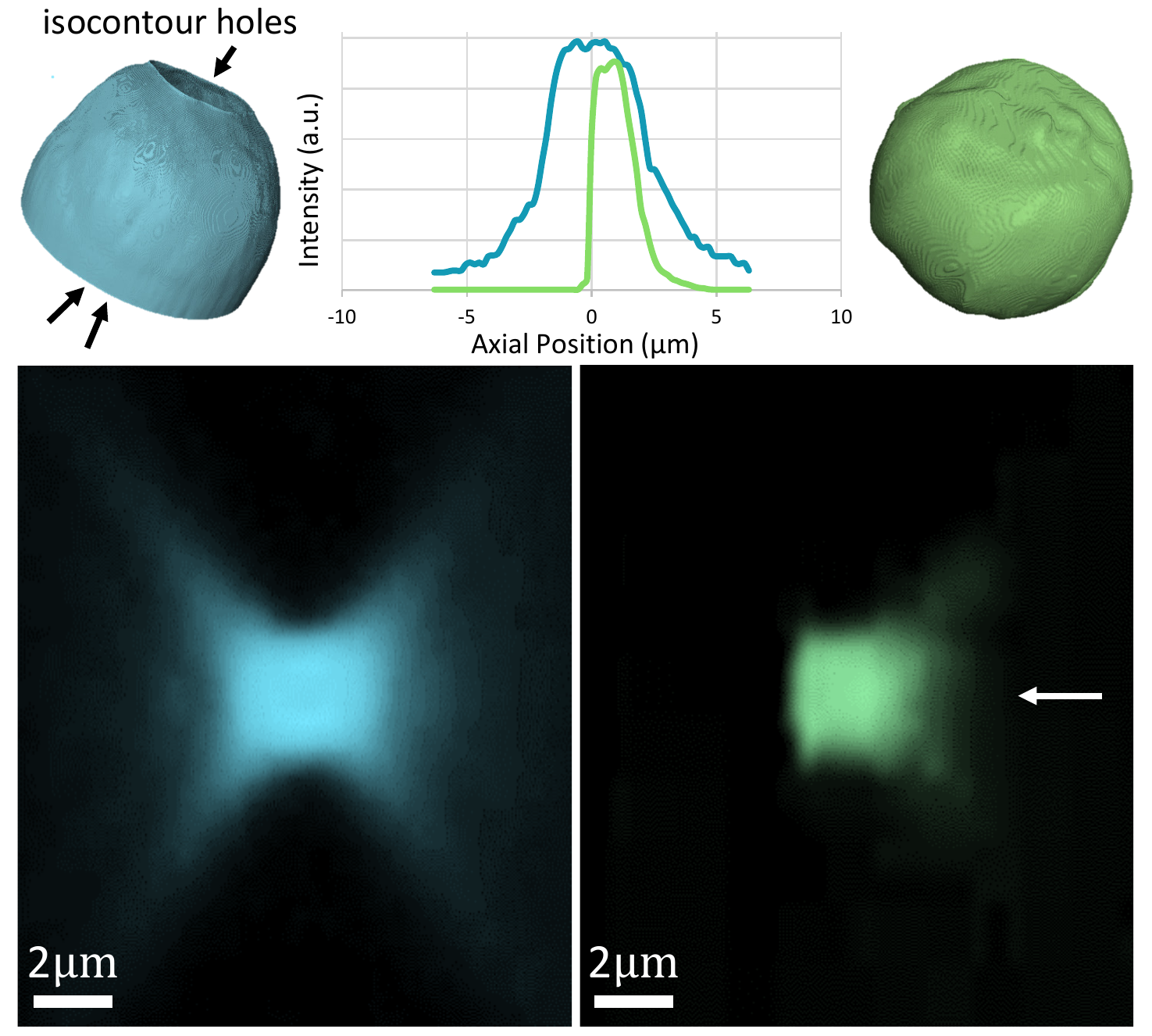}
  \caption{Central profiling of microspheres using a wide-field fluorescence microscope (blue) and MUVE (green). The axial measurements of micro-beads ($\approx$\SI{4}{\micro\metre}) were collected at a \SI{1.0}{\micro\metre} sectioning size for both optical and physical sectioning. Intensity plots were measured across the central line along the cutting direction indicated by a white arrow. 3D volume rendering of large beads ($\approx$\SI{500}{\micro\metre}) showing imaging artifacts such as shadows (indicated by black arrows) involved in optical sectioning microscopy.}
  \label{fgr:CentralProfile}
\end{figure}

The lateral resolution of MUVE is diffraction limited, and similar to fluorescence microscopy is determined by the emission wavelength and objective numerical aperture (NA). We verified the lateral resolution using a USAF 1951 resolution test target (Edmund Optics). Images for this paper were acquired using a 40X Nikon objective (0.6NA). The horizontal construction of our MUVE prototype prohibited the use of immersion objectives, however previous work has already demonstrated MUSE compatibility with water-immersion optics \cite{yoshitake_rapid_2018}.

MUVE axial resolution is dominated by the exponential absorbance of the embedding medium (Figure \ref{fgr:muve_simulations}). The presented prototype provides axial resolution beyond the diffraction limit due to limitations in the NA of air objectives. Further studies will be required to determine the practical PSF in other imaging media.

MUVE resolution benefits were validated by imaging a phantom composed of \numrange{1}{5}\si{\micro\metre} fluorescent green beads (Cospheric Polyethylene Microspheres), (em. \SI{515}{\nano\metre}) which were diluted 1000-fold into UV27-doped paraffin wax. We compared MUVE with wide-field fluorescence microscopy (Nikon Eclipse TI-E Inverted Microscope) using conventional excitation at \SI{390}{\nano\metre} (DAPI excitation). The MUVE axial PSF shows a notable improvement over the traditional pattern of the wide-field microscope (Figure \ref{fgr:CentralProfile}). The benefits of the MUVE PSF come in two forms: (1) physical ablation results in a truncated asymmetric emission spot, since previous layers of the sample have been removed, and (2) absorbance of the doped embedding medium dominates the penetrating half of the PSF. This allows reconstruction of elements (i.e. lower parts of spheres) obstructed using optical sectioning.

\subsection{Block-Face Imaging of Fluorescent Samples}
Mouse organs, including brain, kidney, liver, lung, spleen, and testicle were embedded in UV27-doped paraffin wax and stained after embedding with DAPI. MUVE axial resolution sufficient to resolve individual cell bodies and their chromatin distributions. For example, two types of pneumocytes are distinguishable in the lung image (Figure \ref{fgr:muse}a) and Kupffer cells and hepatocytes are also distinguishable in the liver (Figure \ref{fgr:muse}d). Cerebellar neurons within the granular are also clear to determine, along with their chromatin structure (Figure \ref{fgr:muse}c). The use of oblique UV illumination reveals tissue topographical information with enhanced contrast, consistent with previously published MUSE images \cite{fereidouni_microscopy_2017}. For example, surface profiles of kidney renal tubules are visible using eosin (Figure \ref{fgr:muse}e).

The effectiveness of UV27 doping is shown using conventional microscopy (Figure \ref{fgr:BlockComparison}a,d), MUSE (Figure \ref{fgr:BlockComparison}b,e) and MUVE reconstructions (Figure \ref{fgr:BlockComparison}c,f). Direct comparisons in light penetration and reconstruction behaviors between conventional paraffin embedding, which introduces some UV absorption, and UV27-doped paraffin-embedding tissues. For instance, the 3D reconstruction of paraffin-embedded tissue shows large and rough microvascular structures whereas the 3D reconstruction of UV27-doped paraffin-embedded tissue shows fine and smooth capillaries.

\subsection{Microvascular Imaging}
\begin{figure*}[tb]
  \centering
  \includegraphics[width=0.95\textwidth]{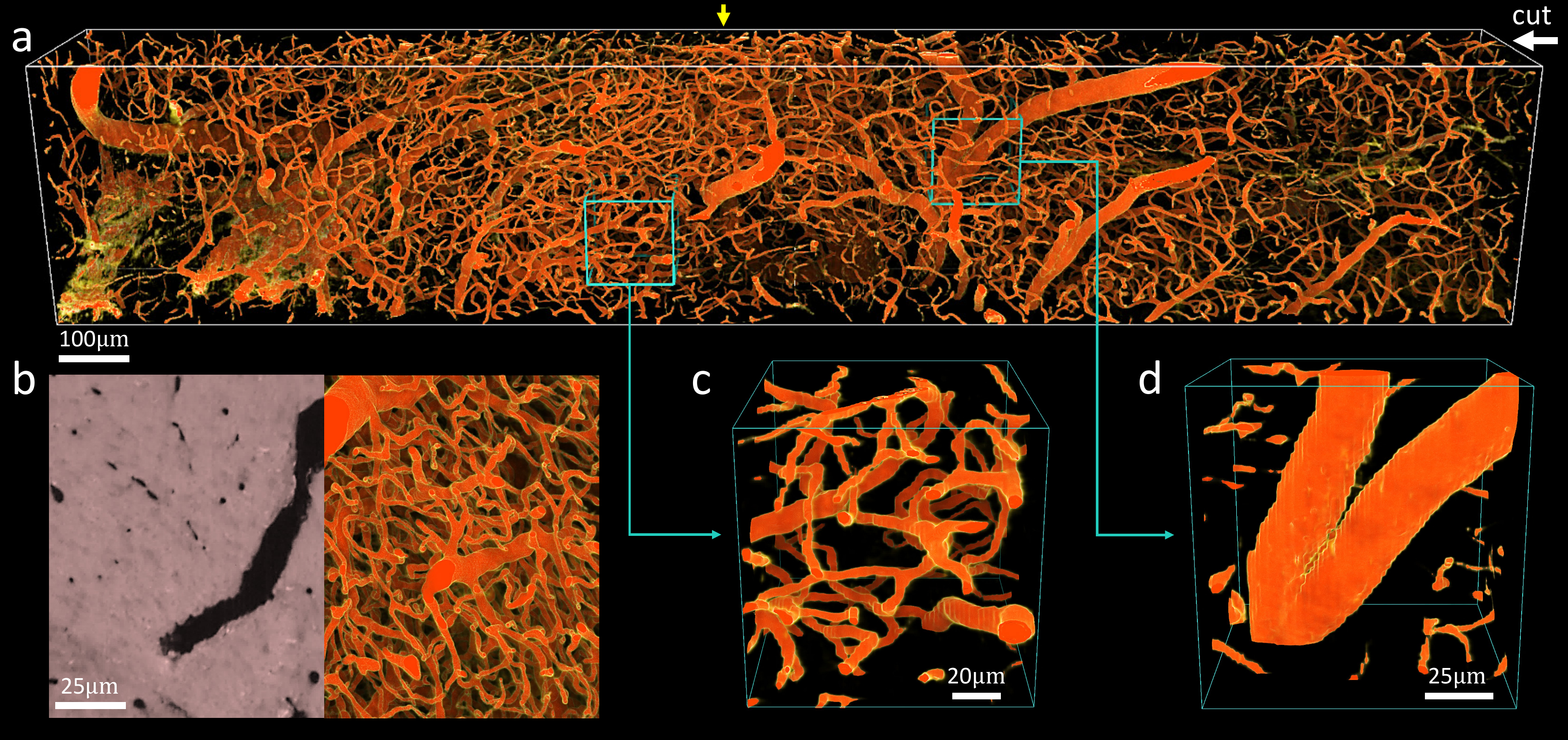}
  \caption{Coronal MUVE imaging of mouse midbrain stained with India-ink. (a) Volume rendering of the entire data set (389$\times$241$\times$2134\si{\micro\meter}) showing the densely-connected microvascular network. (b) One complete cross section (z-axis position indicated by a yellow arrow) with a maximum intensity projection (MIP) overlapped over half. (c-d) Close-up view of small regions (128$\times$128$\times$128\si{\micro\metre}) showing that the sampling resolution of MUVE is large enough to resolve microvessels with different sizes.}
  \label{fgr:DeepNetwork}
\end{figure*}

UV excitation is compatible with absorbing (negative) stains, where contrast is provided by exciting auto-fluorescence in the surrounding tissue and embedding compound. This is particularly useful for microvascular reconstructions using India-ink (Figure \ref{fgr:DeepNetwork}), which mitigates the need for expensive fluorescent alternatives such as lectins \cite{porter1990differential}. This data set was imaged with a lateral sample spacing of \SI{0.37}{\micro\metre} and a \SI{2.0}{\micro\metre} cutting interval. Such spatial resolution is capable of reconstructing the smallest capillaries and their surface profiles.

\begin{figure}[tb]
  \centering
  \includegraphics[width=\the\columnwidth]{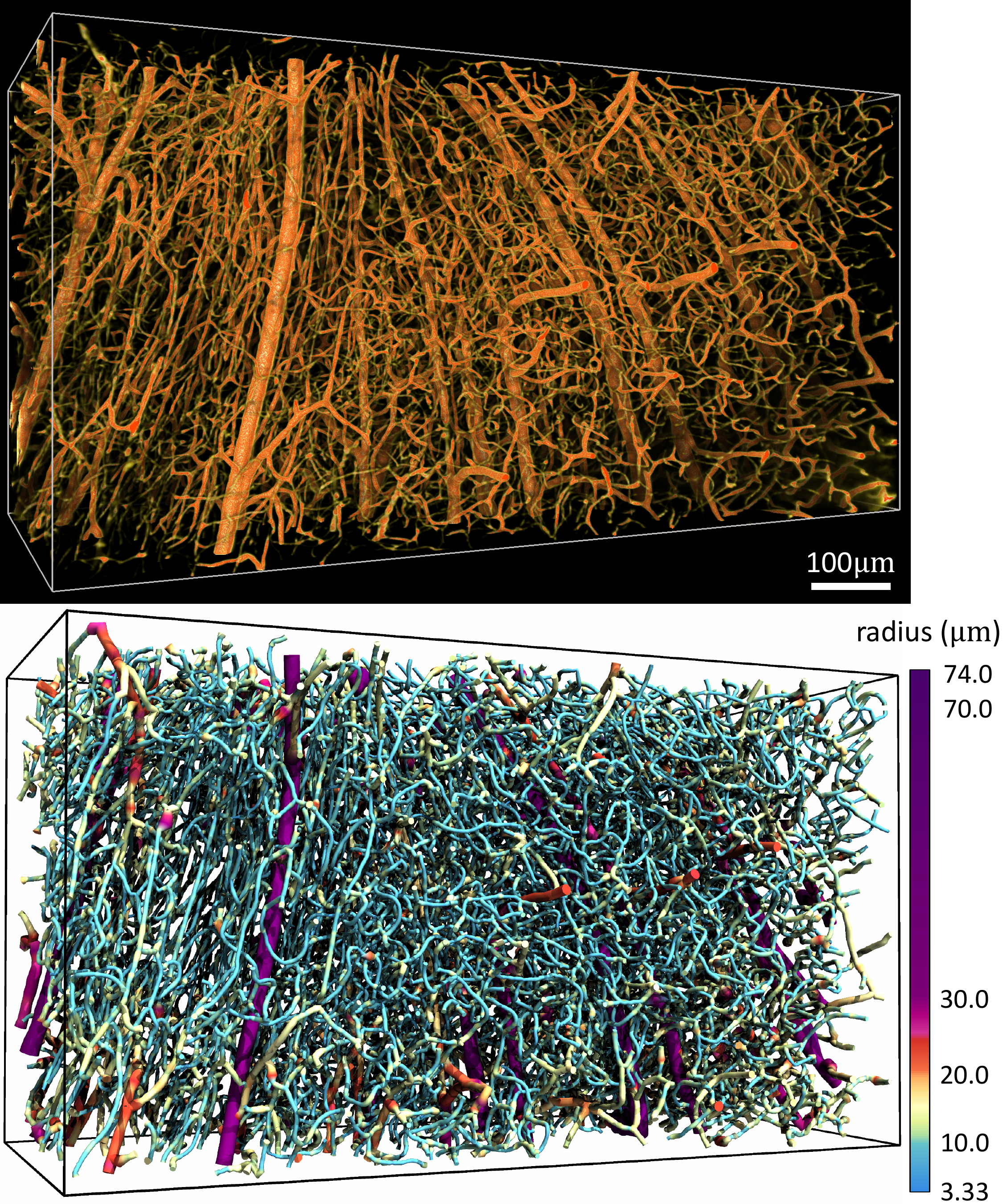}
  \caption{Mouse cerebral cortex microvasculature (top) and segmentation (bottom). The smallest microvessel (rendered in blue) acquired in this data set is $\approx$\SI{6}{\micro\metre} in diameter while the largest (rendered in purple) is $\approx$\SI{148}{\micro\metre}.}
  \label{fgr:Segmentation}
\end{figure}

While the varying vessel thickness can introduce gaps in volume visualizations (Figure \ref{fgr:DeepNetwork}d), the images are particularly high contrast and simple to segment using existing algorithms \cite{govyadinov_robust_2019}. This allowed us to create an explicit graph model with approximately 8,000 edges and 100,000 vertices of a cortical microvascular network (Figure \ref{fgr:Segmentation}), which was visualized using ParaView (Kitware). Note that networks of this size are challenging to reconstruct with optical sectioning due to increased light scattering with sample depth.

\subsection{Microvascular and Nuclear Imaging}
\begin{figure}[!ht]
  \centering
  \includegraphics[width=\the\columnwidth]{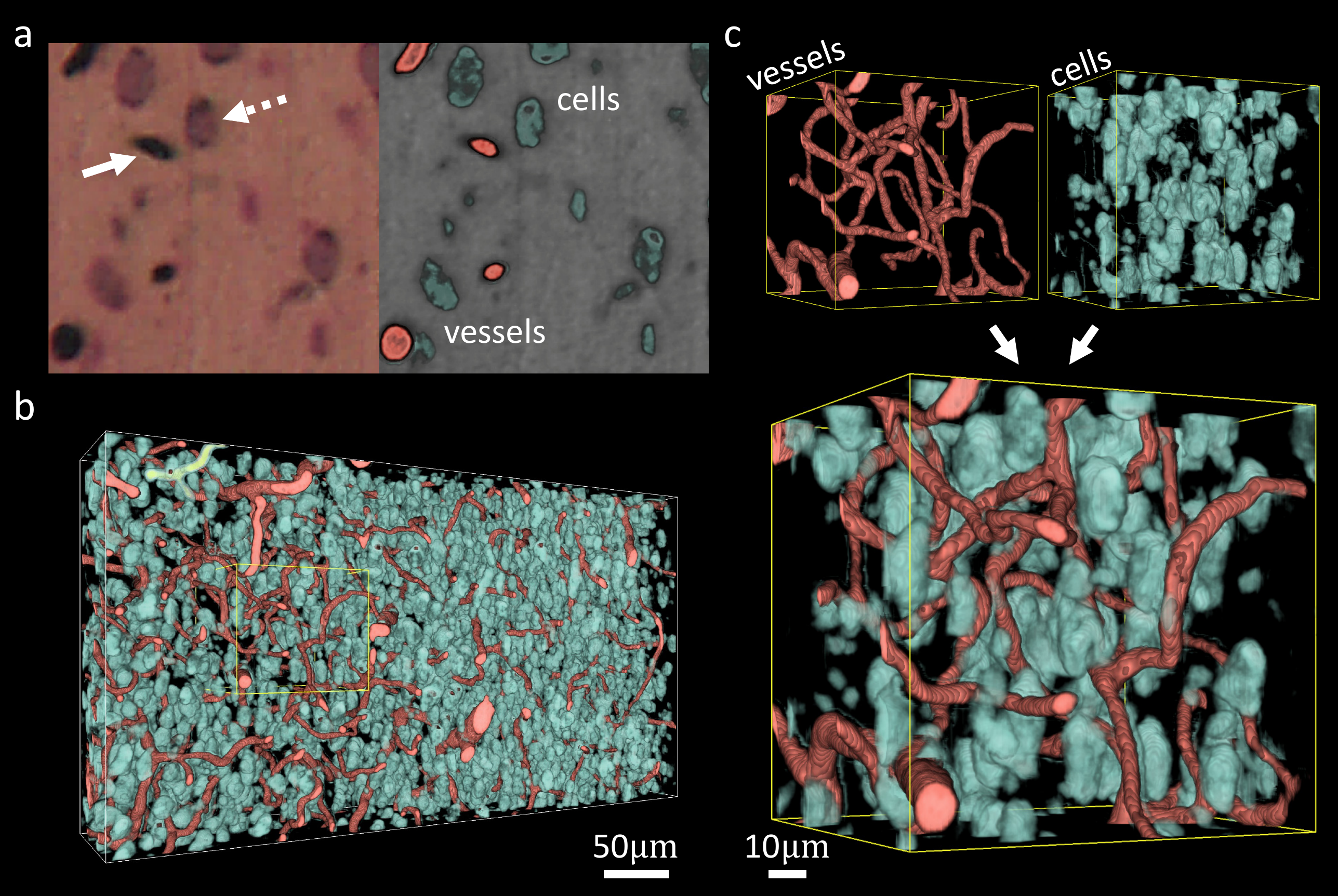}
  \caption{Coronal MUVE imaging of mouse thalamus stained with India-ink and thionin. (a) Tissues are dark red while cell nuclei are dark brown (arrow) and vessels are black (dashed arrow) under UV illumination (left), providing enough contrast to segment both the cellular and vascular structures (right). (b) Volume rendering of the entire data set shows the density and organization of microvasculature with surrounding cellular details. (c) Volume rendering of a small region (100$\times$100$\times$100\si{\micro\meter}) shows series of connected microvessels (red) along with their associated cells (green), and separate channels show detailed cellular and vascular structures.}
  \label{fgr:CellVessel}
\end{figure}

Finally, we investigate combination staining of both microvasculature and nuclei in the brain using thionine with India ink perfusion (Figure \ref{fgr:CellVessel}). A region of the mouse thalamus was imaged at a \SI{0.37}{\micro\metre} lateral resolution with \SI{1.0}{\micro\metre} axial sections to resolve cell nuclei (Figure \ref{fgr:CellVessel}b). This demonstrates potential for studying cellular-vascular relationships linked to many neurodegenerative diseases. This data set was manually segmented using the multi-thresholding function in Amira (ThermoFisher) and visualized using volume rendering (Figure \ref{fgr:CellVessel}c).

\section{Discussion and Future Work}
We have introduced a high-throughput imaging methodology for multiplex imaging of large-scale samples at sub-micrometer resolution at low cost. MUSE milling is capable of imaging densely-interconnected microvascular networks, opening the door to simple acquisition and quantification of capillary changes common during disease progression \cite{castillo_carranza_cerebral_2017} and guide the fabrication of \textit{in vitro} disease models \cite{guo_accurate_2019}. The proposed technique is compatible with a wide range of existing objectives, and can be integrated into immersion-based imaging systems to provide lateral resolution equivalent to existing fluorescence techniques. While MUSE milling eliminates constraints on sample depth, additional work on UV doping of embedding compounds will be necessary to further reduce and quantify axial resolution. Finally, the proposed method provides comparable speed to 2D fluorescence imaging, and was able to produce a deep microvascular network ($\approx$\SI{2}{\milli\metre}) within 2 hours using an automated microtome. While the prototype is limited to a single FOV, custom microtome using 3-axis stages can provide a cost-efficient technology that is simple to build and maintain in most laboratories.

\section*{Acknowledgments} 
\addcontentsline{toc}{section}{\hspace*{-\tocsep}Acknowledgments} 
We would like to thank Pavel Govyadinov at University of Houston for his help on microvascular segmentation. This work was funded in part by the National Science Foundation I/UCRC BRAIN Center \#1650566, Cancer Prevention and Research Institute of Texas (CPRIT) \#RR140013, the University of Houston Division of Research, and the National Institutes of Health (NIH) / National Heart, Lung, and Blood Institute (NHLBI) \#R01HL146745 and National Cancer Institute (NCI) \#1R21CA214299.

\small
\bibliographystyle{naturemag}
\bibliography{bib_muve}


\end{document}